# Discovering similar Twitter accounts using semantics

Gerasimos Razis, Ioannis Anagnostopoulos
Computer Science and Biomedical Informatics Dpt., University of Thessaly
*{janag, razis}@dib.uth.gr*

**Abstract:** On daily basis, millions of Twitter accounts post a vast number of tweets including numerous Twitter entities (mentions, replies, hashtags, photos, URLs). Many of these entities are used in common by many accounts. The more common entities are found in the messages of two different accounts, the more similar, in terms of content or interest, they tend to be. Towards this direction, we introduce a methodology for discovering and suggesting similar Twitter accounts, based entirely on their disseminated content in terms of Twitter entities used. The methodology is based exclusively on semantic representation protocols and related technologies. An ontological schema is also described towards the semantification of the Twitter accounts and their entities.

**Keywords:** Similarity network, social semantics, Twitter entities

## 1. Introduction

Microblogging is a form of Online Social Network (OSN) which attracts millions of users on daily basis. Twitter is one of these microblog services, where its users vary from citizens to political persons and from news agencies to large organizations. Obviously, some users are more influential than others, while many tend to have similar interests. We have created InfluenceTracker[1], a publicly available website where anyone can rate and compare the recent activity and influence of any Twitter account.

The aim of this paper is threefold. Firstly, we propose an improvement over a previous work of us, which is used for calculating the importance and influence of a Twitter account. This improvement incorporates a quality measurement that reflects other users' feel or preference over the examined tweets. Secondly, we propose an ontology and its related semantic mechanisms/technologies which allow us to semantify similarity features (mentions, replies, hashtags, photos, and URLs) as Linked Data. Finally, we propose a methodology for rating the similarity of different Twitter accounts. This methodology is entirely based on the contents of the tweets generated by the accounts, and more specifically based on i) the three basic Twitter entities,( mentions, hashtags and URLs), and ii) the web domains that host the URLs. All the necessary information for the implementation of this methodology was retrieved using exclusively SPARQL queries from the graph generated from our proposed ontology.

---

[1] http://www.influencetracker.com

The remainder of this paper is organized as follows. In the next section, we provide an overview over the related work on semantifying and generating RDF graphs from Twitter data, on semantic modeling and recommendation in Twitter, as well as on measuring influence in Twittersphere. In Section 3, we describe our approach in terms of data semantification and modeling, as well as how we rate the influence of a Twitter account. In Section 4, we analytically present the implemented on-line service and the ontology behind it that transforms raw data from the Twitter API into an RDF graph. Section 5 describes our approach towards similarity recommendation for Twitter accounts. In order to gain insight into this methodology we describe and a case study. In Section 6 we evaluate and discuss the results of the case study, and we further evaluate our methodology against subjective ratings from 22 evaluators. Finally, Section 7 provides the conclusions of our work by summarizing the derived outcomes, while providing considerations on our future directions.

## 2. Related work

This section provides an overview over the related literature on discovering influential users, and on related modeling and recommendation techniques in terms of content-based and data-driven approaches.

### 2.1. Measuring influence in Twitter

The calculation of the impact a user has on social networks, as well as the discovery of influencers in them is not a new topic. It covers a wide range of sciences, ranging from sociology to viral marketing and from oral interactions to Online Social Networks (OSNs). In the related literature, the term "influence" has several meanings and it is differently considered most of the times.

Romero et al. (Romero et al., 2011) utilized a large number of tweets containing at least one URL, their authors and their followers. Their aim is to calculate how influential or passive the Twitter users are. The produced influence metric depends on the "Follower-Followee" relations of the users, as well as their retweeting behavior. The authors state that the number of followers a user has, is a relatively weak predictor of the maximum number of views a URL can achieve. As our work has shown (Razis and Anagnostopoulos, 2014a), the number of followers an account has does not guarantee the maximum diffusion of information in Twitter. This is because, in order to achieve high levels of diffusion, your followers should not only be active, but they should also have a high probability of retweeting, thus transmitting the messages they receive to their followers.

The work described in (Cha et al., 2010) proposes a methodology where for each Twitter user, three different types of influence are introduced. These types are "Indegree" (number of followers), "Retweet" (number of user generated tweets that have been reweeted) and "Mention" (number of times the user is mentioned in other users' tweets). A necessary condition for the computation of these influence types is the existence of at least ten tweets per user. The authors claim that "Retweet" and "Mention" influence correlate well with each other, while the "Indegree" does not. Therefore, they come up with the conclusion that users with high "Indegree" influence are not necessarily influential.

A topic-oriented study on the calculation of influence in OSNs is presented by Weng et al. (Weng et al., 2010). The authors propose an algorithm which takes into consideration both the topical similarity between users and their link structure. It is claimed that due to homophily, which is the tendency of individuals to associate and bond with others having similar interests, most of the "Follower-Followee" relations appear. This work also suggests that the active users are not necessarily influential.

Another approach which defines influence in terms of copying what the directly related account does is presented in (Goyal et al., 2010). In this work, the authors propose an "influenceability" score, which represents how easily a user is influenced by others or by external events. It is built on the hypothesis that a very active user performs actions without getting influenced by anyone. The users of such a type are considered as responsible for the overall information dissemination in the network.

Boyd et al. (Boyd et al., 2010) stated that retweeting can be also characterized as a conversational infrastructure. According to the authors, a conversation "exists" either during a retweet where some new information is added to the initial message, or when a single tweet is retweeted multiple times. The latter is interpreted by the authors as an action to invite new users into the conversation.

All the related studies have shown that the most active users or those with the most followers are not necessarily the most influential. This fact has also been spotted by our work (Razis and Anagnostopoulos, 2014a). As described in Section 3, our Influence Metric depends on several factors, where the account activity is only one of them. Simply put, as the authors in (Srinivasan et al., 2014) state, enormous influence may spring from lesser known persons, while the "celebrities" may not be influencers.

Contrary to the aforementioned studies, for the calculation of our Influence Metric we neither set a lower threshold on the number of the user-generated tweets, nor we only utilize a specific subset of tweets that fulfill certain criteria (e.g. those containing URL etc.). All the Twitter accounts can be used as seed for the calculation of our Influence Metric, thus differentiating our work in respect to the related literature.

## 2.2. Semantic Modeling and Recommendation in Twitter

As semantics and linked data continuously rise, more works relevant to semantic modeling in OSNs appear. The authors in (Celik et al., 2011) and (Abel et al., 2011) propose frameworks for enriching Twitter messages with semantics. The first work involves the identification of semantic relationships between entities by analyzing Twitter posts. These semantic links are between persons, products, events and other entities and are utilized in order to provide suggestion to the users. The latter aims in modeling the users' profiles based on their microblogging activities in order to link Twitter posts with news articles from the Web.

The work presented in (Shinavier, 2010) introduces a semantic data aggregator, which combines a collection of compact formats for structured microblog content with Semantic Web vocabularies. Its main purpose is to provide user-driven Linked Data. The main focus of this work is on microblog posts and specifically on their creators, their content and their associated metadata.

Another framework which utilizes semantic technologies, common vocabularies and Linked Data in order to extract and mine microblogging data regarding scientific events from Twitter is proposed in (De Vocht et al., 2011). The authors attempt to identify persons and organization related to them based on time, location and topic categorization.

Although ontologies and semantic technologies have been used in other works, none of them capture and model such a wide range of information, spanning from the Twitter related characteristics of the accounts to the entities found in the posted messages. In a previous work of ours (Razis and Anagnostopoulos, 2014b) we proposed an ontological schema towards semantification provision of Twitter analytics.

Finally, content-based and data-driven approaches have been used for estimating a Twitter user's location (Cheng et al., 2010), as well as the interestingness in terms of diffusion and the content of the tweets (Naveed et al., 2011). In a previous work of ours (Anagnostopoulos et al., 2015), we utilize the data retrieved from Twitter in order to investigate the query suggestion provision that can be extracted from large graphs, having no prior knowledge of them. Towards this direction, an algorithmic approach is introduced for creating a dynamic query suggestion set which consists of the most viral and trendy Twitter entities (hashtags, mentions and URLs) with respect to a user's provided query input.

Finally, a lot of interest has also been paid in recommendation systems based on those approaches. Buzzer, as described in (Phelan et al., 2011), is a service that suggests news articles to Twitter users, by not only mining terms from their timeline, but also from their friends' timeline. These terms act as ratings for promoting and filtering news content. The authors in (Hannon et al., 2010) also utilize Twitter content in order to capture the interests of the users (and their related ones) in order to recommend them new followers or followees. URLs as a recommendation factor are studied in (Chen et al., 2010) in terms of direct users' attention in more focused information streams.

## 3. Rating Influence in Twitter – Our approach

If depicted in graph, Twitter accounts are represented by nodes. Edges that connect these nodes are the relations of "Follower-Followee" instances. Even if some accounts are more influential than others, the influence measurement does not merely depend on the number of "Followers", even if that number is big enough. In case that the number of "Followees" is larger, then the user could be characterized as a "passive" one. That type of users is regarded as those who are keener on viewing or being informed through tweets rather than composing new ones. Therefore, a suitable factor is the ratio of "Followers to Followees" (*FtF* ratio). Another important factor is the tweets creation rate (*TCR*). For example, let us see the case where two accounts have nearly the same *FtF* ratio. Obviously, the account with the higher *TCR* tends to be more influential. In our methodology, and in order to calculate that rate, we process the latest 100 tweets as provided from the Twitter API. That helps us to keep dynamic the values of *TCR* (and consequently the Influence Metric), as it depends on the most recent activity of the accounts in Twitter. In order to maximize the precision of the metric, the timeframe of its calculation is measured in hours.

Each tweet is associated with several other kinds of information presented in influencetracker.com. Two of them are the "Retweets" and "Favorites" counts, which represent how many times a Tweet has been retweeted as well as marked as favorite by other users respectively. In our methodology, we utilize these counts in order to calculate the h-index of the "Retweets" and "Favorites", over the last 100 tweets of an examined account. The aim of these measurements is to provide a quality overview of the tweets of a Twitter account in terms of likeability and impact in Twittersphere. These indexes are based on the established h-index (Hirsch, 2005) measurement and are named *"ReTweet h-index - Last 100 Tweets"* and *"Favorite h-index - Last 100 Tweets"*. The most important factor regarding them is that they reflect other users' assessment of the content of the tweets.

Consequently, a Twitter account has *"ReTweet h-index - Last 100 Tweets"* equal to *h*, if *h* over the last *Nt* tweets have at least *h* retweets each, and the remaining *(Nt - h)* of these tweets have no more than *h* retweets each (max. *Nt*=100). This can be interpreted as follows: at least *h* tweets have been retweeted at least *h* times. Thus, we consider that this retweeting action results in the generation of at least *h\*h* new tweets, which have to be attributed to the account that initially posted them.

However, prior to incorporating this amount of new tweets into the equation of the Influence Metric, we employ a calculation mechanism for avoiding outliers. Moreover, we introduce a value called "Adjusted Tweets" which is defined in Equation 1.

$$\text{Adjusted Tweets} = a \times 10^b, \text{ where } 0 < a < 100 \text{ and } a \in \mathcal{R} \quad (1)$$

"Adjusted Tweets" are actually a form of expressing the *h\*h* value. Where applicable, "a" is a two-digit number. Then, that number is divided by 10. The resulting quotient is combined with the Order of Magnitude of the *h\*h*, which is represented by "b", thus forming the "Adjusted Tweets Number" according to Equation 1. Some characteristic examples are provided in Table 1.

As already mentioned, the tweets generated from the retweeting process have to be attributed to the account that initially posted them. Therefore, the value of the "Adjusted Tweets" is added to the 100 tweets retrieved from the account, as defined in Equation 2. The *FtF* ratio is placed inside a base-10 log for avoiding outlier values. Moreover, this ratio is added by 1, so as to avoid the metric being equal to 0 in case where the values "Followers" and "Followees" are equal. In Equation 2, OOM stands as the Order-Of–Magnitude of the Followers. For example, if an account has 10.000 followers then OOM equals to 4.

$$\text{Influence Metric} = \frac{\text{tweets}_k + \text{AdjustedTweets}_k}{\text{Hours}_{\text{since } k_{th}\text{tweet}}} * $$
$$* \text{OOM}(\text{Followers}) * \log_{10}\left(\frac{\text{Followers}}{\text{Followees}} + 1\right), \quad (2)$$

## 4. The Influence Tracker publicly available service

In this section, we present the architecture and infrastructure of influencetracker.com service. In addition, we describe the ontology used for transforming the Twitter feature accounts and the entities included in their tweets (mentions, replies, hashtags, photos, URLs) into an RDF graph.

### 4.1 Architecture

The architecture of the influencetracker.com service and the relevant data flows are presented in Figure 1. The service combines the use of a relational database joint with an RDF triple store. Thus, data and related information displayed at the web pages combine both technologies. The relational database is a MySQL Server and the RDF triple store is contained in an Open Link Virtuoso (OLV) Server. There are two processes of the service.

The first process involves the update of the RDF graph. An implemented service based on Python libraries, is executed on a weekly basis. The process is split into four phases. During the first phase, a request is sent to the Twitter API for each account found in the database. The response contains the data in JSON format. In the second phase, the necessary data are parsed and the metrics are calculated. The third phase involves the semantification of the gathered data with concepts (resources and property URIs) derived from our ontology (see Figure 2) and the RDF graph updates. This process is performed by using the RDFLib framework. During the last phase, the triples are stored in the OLV environment, while the user can use a SPARQL endpoint[2,3] for semantic search.

The second process is a subset of the previous one. It takes place when a Twitter account is searched through the provided web interface. Another service, also implemented in Python, performs a request to the Twitter API for the investigated account. A response is returned in JSON format. In case of a valid account, the necessary data are parsed, related metrics are calculated and stored in the relational database. In this use case, no data are stored at the RDF graph. This is because we wanted to maximize the responsiveness of our service, minimizing in parallel the execution time. Finally, in case where a new account is inserted into our system, the necessary data will be stored at the RDF graph during the next update process.

### 4.2 The InfluenceTracker Ontology

Our ontology utilizes properties from the FOAF ontology (Brickley and Miller, 2014). FOAF (Friend-of-a-Friend) is an ontology for describing persons, their activities as well as their relations to other people and objects, while it can be generalized as to describe all type of entities, called agents, who are responsible for specific actions (Brickley and Miller, 2014). In our context the agents are the Twitter users, who are responsible for specific actions, such as owing Twitter accounts, posting tweets, interacting with others etc. Figure 2 displays the classes and their hierarchical relationships. Highlighted are the FOAF ontology classes.

During the representation of the entities, two specific prefixes are used, namely "foaf" and "it". They correspond respectively to the namespace of the FOAF and of our proposed ontology. The ontology is built on three basic building blocks, namely classes, as well as object and datatype properties.

*4.2.1. Classes*

The classes are used to represent conceptual entities. Those defined in the influenceTracker ontology are the following:

---

[2] http://www.influencetracker.com/endpoint
[3] http://www.influencetracker.com:8890/sparql

- foaf:Agent: A general class which describes agents who are responsible for several actions (Brickley and Miller, 2014).
- it:User: It is a subclass of the foaf:Agent and describes the agents that own a Twitter account. These may be physical persons, organizations, events, parties etc.
- foaf:OnlineAccount: It represents the provision of some form of online service, by some party (indicated indirectly via the foaf:accountServiceHomepage object property) to some foaf:Agent (Brickley and Miller, 2014).
- it:TwitterAccount: The class is a subclass of the foaf:OnlineAccount and represents the actual Twitter accounts.
- it:GeneralInfo: The class contains the Twitter related details of an account characterized by influencetracker.com as "General Information". These are the total number of tweets, the *TCR*, the retweet ratio, and the number of followers and following.
- it:QualityMetrics: The class contains the metrics of a Twitter account characterized by influencetracker.com as "Quality Metrics". These are the *"ReTweet and Favorite h-index - Last 100 Tweets"*, the estimated *"ReTweet and Favorite h-index"*, the reply ratio and the value of our influence metric.
- foaf:Document: The class represents those things which are, broadly conceived, documents. There is no distinction between physical and electronic ones (Brickley and Miller, 2014).
- foaf:Image: The class corresponds to those documents which are images. It is a subclass of the foaf:Document, since all images are documents. Digital images are instances of this class (Brickley and Miller, 2014).
- it:Hashtag: The class describes the entities which are hashtags (words starting with "#").
- it:URL: The class describes the entities which are URLs.

*4.2.2. Object Properties*

The object properties are those for which their value is an individual. Those defined in the influenceTracker ontology along with their concept restrictions are the following:
- foaf:account: The property is used to relate a foaf:Agent to a foaf:OnlineAccount for which they are the sole account holder (Brickley and Miller, 2014).
- foaf:accountServiceHomepage: The property indicates a relationship between a foaf:OnlineAccount and the homepage of the supporting service provider (Brickley and Miller, 2014).

- it:hasGeneralInfo: The property relates an it:User to an it:GeneralInfo which contains the Twitter related information of the owned account, characterized by influencetracker.com as "General Information".
- it:hasMentioned: The property relates an it:User to an it:User that has been mentioned in the first user's tweets.
- it:hasQualityMetrics: The property relates an it:User to an it:QualityMetrics which contains the metrics of the owned account, characterized by influencetracker.com as "Quality Metrics".
- it:hasRepliedTo: The property relates an it:User to an it:User that has received a tweet as a reply from the first user.
- it:includedHashtag: The property relates an it:User to an it:Hashtag that has been included in the user's tweets.
- it:includedImage: The property relates an it:User to an it:Image that has been included in the user's tweets.
- it:includedUrl: The property relates an it:User to an it:URL that has been included in the user's tweets.
- it:isFollowing: The property relates an it:TwitterAccount to an it:TwitterAccount in cases where the first account follows the second one. It represents the action named "Follow" introduced by Twitter. It is the reverse property of it:hasFollower.
- it:hasFollower: This property relates an it:TwitterAccount to an it:TwitterAccount in cases where the second account is a follower of the first one. It is the reverse property of it:isFollowing.

These properties have been defined in such a way so as to be easily extensible to cover concepts from other OSNs as well. The Twitter accounts can be replaced by those of Facebook and the tweets by the statuses. The actions of "Share" and "Like" found in Facebook are the equivalent of "Reteweet" and "Favorite" of Twitter. The concepts of hashtags, mentions, replies, images and URLs are the same in these OSNs.

*4.2.3. Datatype Properties*

The datatype properties are those for which their value is a data literal. Those defined in the influenceTracker ontology along with their concept restrictions are the following:
- foaf:accountName: The property provides a textual representation of the account name (unique ID) associated with that account (Brickley and Miller, 2014).
- it:description: The property provides the description of an account, as set by its owner.
- it:displayName: The property provides the name displayed at the web page of an account, as set by its owner.

- it:followers: The property provides the number of the followers of an account.
- it:following: The property provides the number of the accounts that an account follows.
- it:hIndexFav: The property provides the value of the *"Favorite h-index - Last 100 Tweets"* metric of an account.
- it:hIndexFavDaily: The property provides the estimated daily value of the *"Favorite h-index"* metric during the lifespan of an account.
- it:hIndexRt: The property provides the value of the *"ReTweet h-index - Last 100 Tweets"* metric of an account.
- it:hIndexRtDaily: The property provides the estimated daily value of the *"ReTweet h-index"* metric during the lifespan of an account.
- it:imageUrl: The property provides the URL that leads to an image which was included in a tweet.
- it:influenceMetric: The property provides the value of the Influence Metric measurement. Its aim is to describe both the importance and impact of an account in a social network.
- it:profileLocked: The property indicates whether the profile of an account is publicly visible or not.
- it:activeAccount: The property indicates whether the an account is active or not.
- it:replyRatio: The property provides the ratio of the user's latest tweets which are used as replies to other users' tweets.
- it:retrievedOn: The property provides the date that the information regarding an account was lastly updated.
- it:rtPercent: The property provides the percentage of the latest user's tweets that are retweets from other accounts.
- it:tweets: The property provides the number of the total tweets posted by an account.
- it:tweetsPerDay: The property provides the number of the average tweets posted per day by an account.
- it:url: The property provides the short URL that leads to a web site which was included in a tweet.
- it:fullUrl: The property provides the full URL representation of a shortened one which was included in a tweet.
- it:domain: The property provides the domain of a URL which was included in a tweet.

As already mentioned, a public endpoint allows the search of the collected semantic data. The URIs which are returned by the queries are dereferenceable ones, consequently the resources that they identify are represented by documents, which in our case are in HTML format. Specifically, these URIs are constructed using the Slash format. An example of such URI is "www.influencetracker.com/resource/User/youtube". It represents the document where the resource "youtube", an instance of the "it:User" class, is described. An instance of that document can be found at Figure 3.

## 5. Similarity Recommendation in Twitter

As already mentioned the Twitter accounts and their related information in tweets, named entities, (e.g. mentions, replies, URLs, hashtags, photographs) are retrieved and stored in an RDF graph. Obviously, many of these entities are found in many tweets posted by different Twitter accounts. The more entities the accounts have used in common, the more similar their content tends to become. The methodology of calculating the similarity of Twitter accounts is presented below.

As suggested in (Naveed et al., 2011) the presence of hashtags, mentions and URLs is typical in a tweet and were utilized in their content-based framework. For the calculation of our Similarity Metric, four entities are used as comparison coefficients: the three "typical" ones (i.e. hashtags, mentions, and URLs) and additionally the domains of those URLs that an account has included in its tweets. The proposed methodology consists of the following seven steps:

1. define *k*, that is the number of the top similar accounts to be discovered,
2. define the *depth* of the similar accounts to be discovered (e.g. if depth equals to two the top-k similar accounts of the top-k ones of the examined account will be searched and so forth),
3. define the account to discover its top-k similar ones,
4. retrieve the entities of an entity category of the examined account (e.g. all the hashtags included in its tweets),
5. discover all Twitter accounts that included those entities in their tweets,
6. for each Twitter account find their total entity counter of the specific category $E_N$, (i.e. how many hashtags have been tweeted),
7. find the common amount of the specific category ($E_{CN}$) between the examined account in respect to others (e.g. their common amount of hashtags),
8. calculate the coefficient of that specific category of Twitter entity ($E_{Cf}$) between the examined and each one of the other accounts (e.g. hashtag coefficient),
9. repeat steps 4 to 8 for the remaining entity categories,
10. depending on the depth value repeat steps 3 to 9.

The coefficient of a specific entity category ($E_{Cf}$) is defined as the division of the common amount (counter) of the category ($E_{CN}$) by its total entity counter ($E_N$). The calculation of this coefficient is presented in Equation 3. After step 7, four coefficients are calculated, namely mention, hashtag, URL and domain coefficient.

$$E_{Cf} = \frac{E_{CN}}{E_N}, \text{ where: } 0 \leq E_{Cf} \leq 1,\ E_{Cf} \in \mathcal{R},\ E_N > 0 \qquad (3)$$

The next step is to utilize the resulting coefficients in order to calculate the Similarity Metric. Apart from the aforementioned coefficients, there are three other factors that are considered for the calculation of the Similarity Metric.

The first is the frequency of use of each of the four entity categories by the examined user, named "Entity Weight" ($E_W$). It is defined as the division of the entity counter of a specific entity category ($E_N$) by the sum of the entity counters of all entity categories ($E_{SN}$) and it is defined by Equation 4. Four weights are calculated -one for each entity category- namely mention, hashtag, URL and domain weight.

$$E_W = \frac{E_N}{E_{SN}}, \text{ where: } 0 \leq E_W \leq 1,\ E_W \in \mathcal{R},\ E_{SN} > 0 \qquad (4)$$

This factor is useful in cases where the $E_{Cf}$ of an entity category of a compared account is high and the $E_W$ of the examined account is significantly low. Moreover, the $E_W$ is used for properly adjusting outlier $E_{Cf}$ values. The resulting Weighted Coefficient (WC) of a specific entity category is defined in Equation 5.

$$E_{WC} = E_{Cf} * E_W \qquad (5)$$

However, there are cases where the $E_{WC}$ coefficient is not enough. This is the case where two users have the same $E_{Cf}$ value for an entity category, but different $E_{CN}$. The Twitter account with the largest number of $E_{CN}$ is regarded as more similar in respect to the examined account. Therefore, another factor taken into consideration is the number of the intersected (common) entities $E_{CN}$. The resulting Common Coefficient (CC) of that entity category is calculated as presented in Equation 6.

$$E_{CC} = E_{Cf} * E_{CN} \qquad (6)$$

By combining the two aforementioned factors into one equation, we calculate the Common Weighted Coefficient (CWC) of an entity category (Equation 7).

$$E_{CWC} = E_{Cf} * E_{CN} * E_W \qquad (7)$$

The third factor that should be considered before calculating the Similarity Metric is the number of entity categories that the compared account has at least one entity in common with the examined account (label). This factor is used in order to adjust the metric by considering the existence of the number of the four distinct coefficients. Finally, the Similarity Metric (SM) is calculated by incorporating the four coefficients and the three factors into Equation 8.

$$\text{Similarity Metric} = (\text{hashtag}_{CWC} + \text{mention}_{CWC} + \text{URL}_{CWC} + \text{domain}_{CWC}) * \frac{\text{label}}{4}$$

$$\text{where: label} = \{0, 1, 2, 3, 4\}, 0 \leq SM, SM \in \mathcal{R} \qquad (8)$$

All of the aforementioned coefficients and factors are based on the individual characteristics of each Twitter account, thus forming a dynamic and unique Similarity Metric for each pair of examined - compared Twitter account.

### 5.1 Case study

As a case study scenario we applied our proposed methodology on the Twitter account of the ex-minister and current member of the Greek parliament @adonisgeorgiadi. We selected this account since it is well known, highly influential and active. We explicitly claim that we use this account for research purposes and we are not against or in favor in respect to its disseminated content. The aim of this case study is to discover its top-k similar accounts where *k*=15.

The resulting dataset is a graph which was queried in order to apply the proposed methodology in RDF format and it is publicly available in this link[4]. The data can be queried through the provided endpoint of infleuncetracker.com service under the named graph http://influenceTracker/twitterGraph/full. The information in respect to the case study was collected between Oct 13 and Oct 22 of 2014. A quick overview of the contents can be found in Table 2. The graph contains 90,578 Twitter accounts. All the information described in our ontology has been modeled for 2,423 of them[5]. The latter were randomly selected from the mentions found in the captured tweets. It should be noticed that influencetracker.com is an active site, therefore the reader of this article may find also additional accounts (new accounts added after this submission). The remaining 88,155 accounts are followers or are being followed by the fully modeled 2,423 accounts. In addition, there are also 188,542 shortened

---
[4] https://www.dropbox.com/s/i1ow3jt2dgdxzhn/itGraphFull.rar?dl=0
[5] http://www.influencetracker.com/searchedAccounts

URLs, while 72,931 among them have been transformed from tiny to typical URLs in order to retrieve their domains. For this operation we used the http://unshorten.it/ service. The transformed URLs are hosted by 8,402 unique domains. Finally, 38,020 hashtags and 59,160 images are modeled as these were contained in the captured tweets. All the presented data are modeled in nearly 2 million triples.

### 5.2 Case study results (depth=1)

The top-15 similar accounts of @adonisgeorgiadi according to our methodology are illustrated in Figure 4. The nodes correspond to Twitter accounts, while the curving edges indicate a clockwise direction from the source node (@adonisgeorgiadi) to the target node. The thicker the edges the more similar we consider the connected nodes. The edges have the same color as their destination node. The presented network –as well as the others below- is created with the open graph visualization tool named Gephi (layout type: Yifan Hu).

We noticed that 12 out of the top-15 connected accounts, belong to current members of the Greek parliament, while the remaining three accounts belong to a well known political journalist in Greece (@nchatzinikolaou) and to two persons (@app_117, @iptamenos23) who are posting tweets about the political situation in Greece and retweet messages of many politicians.

The first column of Table 3 presents these top-15 similar accounts, while the rest columns highlight the respective factors and metrics defined and presented previously. As can be clearly seen, the presence of each distinct metric (and its respective value) affects the final Similarity Metric (SM) that is depicted in the final column of Table 3. For example, the account @evangantonaros is ranked as the third highest account according to CWC. Nevertheless, due to having one category less in common with the examined account (no common URLs are found), the "label" parameter of Equation 4 equals to 3, thus reducing the Similarity Metric by 33.3%. Finally, @evangantonaros account is ranked as the sixth more similar account among the top-15.

### 5.3 Case study findings (depth=2)

As an extension of the previous experimentation, our next step was to increase the value of *depth* in order to discover the top-15 similar accounts of those displayed in Figure 4. The proposed methodology was implemented iteratively for each one of the previous accounts. The produced network is illustrated in Figure 5.

Moreover, the resulting network consists of 107 Twitter accounts. Approximately the two third of them belong to current members of the Greek parliament, as well as to persons who are actively engaged with political parties in Greece or even official political party accounts. The rest accounts belong to journalists and to persons posting Tweets about the political situation in Greece.

Those 107 nodes representing Twitter accounts are interconnected through 240 directed and weighted edges. Many of the accounts contained in the top-15 similar results are repeated, and therefore the actual number of appeared accounts is less than the maximum that can be achieved. In the presented scenario where depth=2 the maximum number of unique nodes would be 241 (that is 16 sets of top-15 similar accounts plus @adonisgeorgiadi). It is worth noticing that the whole network is built on the 44.6% (107) of the maximum possible nodes (241), since the rest 134 appear again among the top-15 similar accounts one depth further.

An example of an account being in other top-15 results is the examined-root account @adonisgeorgiadi. It appeared in the top-15 results of 13 similar accounts, thus its In-Degree$_{top-15}$ is 13. In a sense, this mutual similarity defines that the connected nodes are highly possible to be similar with at least a 86.67% probability value.

## 6. Evaluation and Discussion

The purpose of this section is two-fold. We first want to evaluate the results of the case study described in the previous section. Then, in order to further evaluate our similarity metrics, we describe a generic evaluation, which involves subjective user ratings for the results obtained from our Similarity Metric.

### 6.1 Case study evaluation

In the previous section, an implementation of the proposed framework for discovering similar Twitter accounts was presented along with some results regarding that use case. In order to verify the validity of the results, our framework was extensively applied on a number of Twitter accounts.

The extended experiment has shown that on average the In-Degree$_{top-15}$ of a "root" account is almost equal to 12. That suggests that there is 80% probability (12 out of 15) of the "inverse" similarity relation to exist between the examined account and its top-15 similar ones. Simply put, if an account B is in the top-15 similar ones of the examined A then there is 80% probability of A being in the top-15 similar accounts of B. This fact reflects the dynamic nature of our Similarity Metric. It is based on the individual characteristics of each account, thus being almost unique for each pair of examined - compared Twitter account.

As already mentioned, when expanding the depth in our network, the theoretical maximum number of nodes is not reached, mainly because mutual similarities between nodes.

Specifically, as the values of *k* and *depth* are increased, the number of the unique nodes in the network is rapidly decreased, while the total number of nodes is increased at a lower rate.

Moreover, Figure 6 presents the theoretical maximum number of accounts (per depth) vs. the actual unique ones inserted into the network. The horizontal axis represents the depth in respect to the initial node (root), while the vertical axis represents the number of examined accounts. The diagram depicts two dotted lines, along with their trend lines of exponential type. The blue dotted line represents the theoretical maximum number of accounts that needs exploring, while the red dotted one the number of unique accounts that eventually explored into the network. Table 4 presents all respective values according to the depth in respect to the root node.

Finally, we also noticed that the more top-k similar accounts are examine the less unique accounts are discovered, as well as the more closed walks (cycles) are found in the generated network. A cycle is a unique closed walk (across different nodes) that starts and ends from a distinct node. In our case study, a cycle of length 5 that starts and ends from the @adonisgeorgiadi node is {@adonisgeorgiadi → @thanosplevris → @vozemberg → @vkikilias → @aris_spiliotop → @adonisgeorgiadi}. Such a cycle reveals a community of similar accounts (all accounts are politician of the same or adjacent political parties).

It's worth noting that such kind of communities follow a power law distribution. Figure 7 presents the cycle (community of similar users) distribution of a network that consists of 365 unique accounts, which were discovered after the methodology was applied for depth=3 and k=15 (sum of last column of Table 4). Vertical axis represents the amount of closed walks, while the horizontal axis represents their size. Among 365 accounts, a total of 531 cycles were revealed. Figure 7 depicts the respective power law distribution.

Two of them have length equal to 68, which is approximately 19% of the total nodes. As the number of the top-k accounts is increased, the average number of nodes per cycle is also increased. Figure 8 depicts the average nodes per cycle distribution of the resulting similarity networks after each depth. Vertical axis represents the amount of the average nodes per cycle, while the horizontal axis the depth. In the presented case the average weighted cycle size is approximately 24.

**6.2 Evaluation against user ratings**

One of the functionalities offered by Twitter to its users is the recommendation of other accounts (as similar) to be followed[6]. These suggestions are personally provided to the users and are mainly based on the users' contacts, e-mail, location, followers and followees, as well as on other public profile information. Very little attention has been given to the content itself (e.g. text or Twitter entities). Moreover, these suggestions are only visible to the account owners and cannot be retrieved using the Twitter API. As a result, it was obvious that we could not evaluate our methodology having as ground truth the respective recommendations provided by Twitter. Thus, in order to further evaluate our methodology, we describe here a generic evaluation over subjective user ratings.

---

[6] https://support.twitter.com/articles/227220-how-to-use-twitter-s-suggestions-for-who-to-follow

Moreover, for the purposes of this evaluation, 22 postgraduate students from an MSc course class at University of Thessaly were engaged. Their task was to subjectively rate the similarity results provided by our methodology. Each student was asked to select an initial root node and then evaluate the similarity network derived when seeking the top-5 similar accounts when the depth search equals to 3. Each individual had to explicitly rate how similar two accounts are –for all separate cases in the resulting network- under a five-point Likert scale, as indicated below:

1. Strongly disagree (totally unsimilar accounts)
2. Disagree (rather not similar accounts)
3. Neither agree nor disagree (I cannot judge – neutral)
4. Agree (the accounts tend to be similar)
5. Strongly agree (I am sure. These accounts are similar)

In other words, 22 distinct case studies were evaluated by every individual, in the same sense as the case study described in sub-Sections 5.1 and 5.2. However, we set the search depth equal to 3, while in order to keep the amount of possible ratings between nodes in fairly levels, we reduced the top-k examined accounts setting k equal to 5.

At Figure 9, we can see the points that indicate the average rating of the evaluators between nodes and according to their distance in the resulting similarity network. As distance, we define the number of hops between the compared nodes. We noticed that when the distance between compared nodes increases, the average subjective similarity rate value (in the five-point Likert scale) decreases. However, this was somehow expected since in a resulting network of the top-k similar accounts of the top-k similar accounts and so forth (according to the selected search depth), the higher similarity values between nodes tend to appear in nodes with lower distances. We also noticed that for low distance values (up to 2) the mean ratings are above 4, denoting that our Similarity Metric works efficiently enough according to the evaluators' opinion.

## 7. Conclusions and Future work

This paper deals with three major areas. Firstly, we revisit a methodology for rating the influence in a given Twitter account. The methodology incorporates an *h*-index-based measurement, reflecting users' actions and preferences over tweets.

Secondly, we introduce an ontology (InfluenceTracker ontology) for semantifying Twitter entities (mentions, replies, hashtags, photos, and URLs) and account characteristics. Information is inserted into an RDF graph, which is publicly available for querying through a provided SPARQL endpoint. To the best of our knowledge, there is currently no active service for providing such kind of data.

Finally, we describe a framework for discovering similar Twitter accounts by exploiting their entities and their relation with other accounts in Twitersphere. For defining the similarity metrics we employ exclusively semantic technologies and models (e.g. see SPARQL Query 1 and 2 in the Appendix) based on the proposed InfluenceTracker ontology. The existence of an ontological scheme and the use of semantics technologies reduce the complexity of storing and retrieving specific segments of data and decrease the number of the necessary calculations required for the computation of the coefficients and metrics presented in Section 5.

The properties found in the InfluenceTracker ontology have been defined and created in such a way so as to be easily extensible to cover concepts from other OSNs as well. For example, Twitter accounts can be replaced by those of Facebook, while the tweets by the statuses. In our methodology, the Facebook actions "Share" and "Like" can be considered as equivalent to "Retweet" and "Favorite" (in Twitter), while concepts of hashtags, mentions, replies, images and URLs are practically identical in these OSNs.

In order to better explain our methodology, we described a case study over a well-known, highly reputable and active Twitter account (@adonisgeorgiadi). We discovered its top-15 similar accounts (depth 1), and then we have further extended our search for the top-15 similar accounts of the previously found ones (depth 2). Finally, we conducted a generic evaluation, which involves subjective user ratings from 22 different evaluators. Results derived that the majority of subjective rates (in average values) were very satisfying.

Our proposal targets to suggest to Twitter users what are the most similar accounts according to their common disseminated content and their relations in the graph. Suggestions offered by Twitter are totally different in concept since they are based on the users' contacts, e-mail, geo-location, and on other public information[7]. Very little attention has been paid to the content of the tweets. In our case, the only factor is the content of the messages (mentions, hashtags, URLs, and domains). Due to our ontological scheme, the content and the relations of an account (followers - followees) are modeled, and therefore only the related accounts would be displayed as the final suggestions.

One of the investigated entities is the domain of the URLs. The more domains the accounts have in common, the higher that coefficient becomes. There are cases where two different domains belong to the same thematic category (e.g. news, sports, science etc.). In the future, we plan to add the thematic category at the most tweeted domains, in order to not only consider "interest" in the same domain, but also in the same thematic domain.

---

[7] https://support.twitter.com/articles/227220-how-to-use-twitter-s-suggestions-for-who-to-follow

Finally, we plan to extend this methodology for highlighting communities (of different sizes) of Twitter accounts of similar content, influence and activities. It's also worth investigating the dynamics of such communities across different thematic domain and real-time events.

# Appendix

```
PREFIX it: <http://www.influencetracker.com/ontology#>
SELECT (COUNT(?ht) AS ?commonHtCounter)
FROM <http://influenceTracker/twitterGraph/full>
WHERE {
<http://www.influencetracker.com/resource/User/{examinedUsername}> it:includedHashtag ?ht .
<http://www.influencetracker.com/resource/User/{randomUsername}> it:includedHashtag ?ht .
}
```

Query 1: Returns the common hashtag counter between two specific users owing Twitter accounts. In our case the first of them is the examined user whose similar accounts will be discovered, while the second one is an arbitrary user to be compared.

```
PREFIX it: <http://www.influencetracker.com/ontology#>
SELECT (COUNT(DISTINCT ?domain) AS ?commonDomainsCounter)
FROM <http://influenceTracker/twitterGraph/full>
WHERE {
<http://www.influencetracker.com/resource/User/{examinedUsername}> it:includedUrl ?urlExamined .
?urlExamined it:domain ?domain .
<http://www.influencetracker.com/resource/User/{randomUsername}> it:includedUrl ?urlUser .
?urlUser it:domain ?domain .
}
```

Query 2: Returns the common domain counter between two specific users owing Twitter accounts. In our case the first of them is the examined user whose similar accounts will be discovered, while the second one is an arbitrary user to be compared.

# Table of Figures

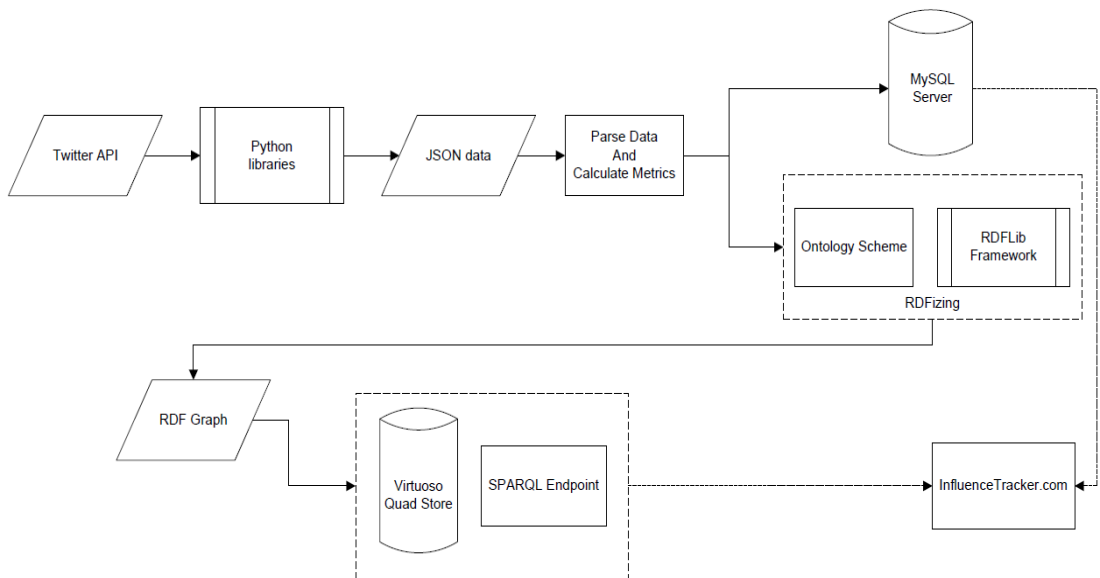

Figure 1: The phases during the updating process of the RDF graph

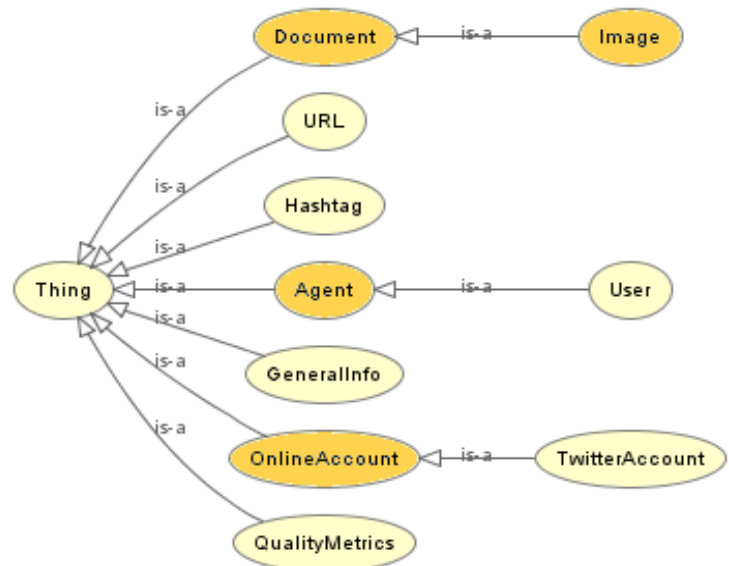
Figure 2: The hierarchy of the classes of the "InfluenceTracker" ontology

Figure 3: The document in HTML format of a dereferenceable URI

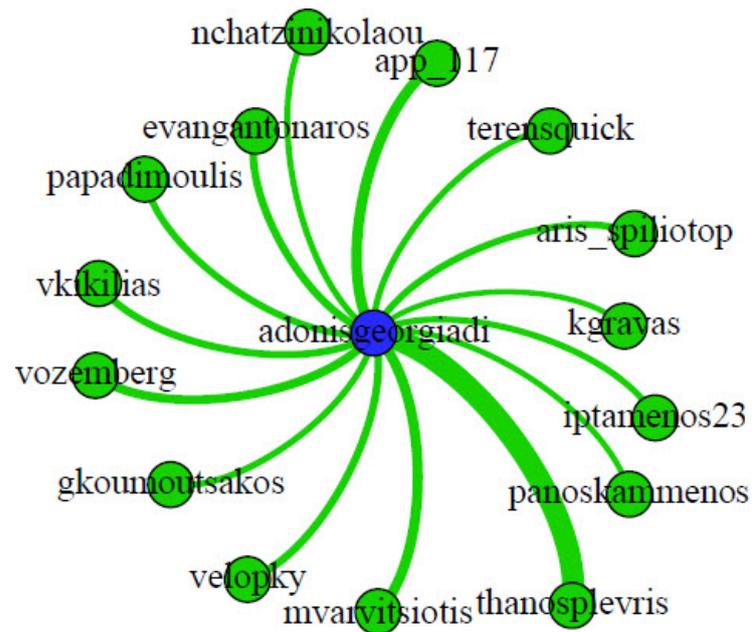

Figure 4: Case study similarity network (depth=1) – Thicker edges denote more similar Twitter accounts

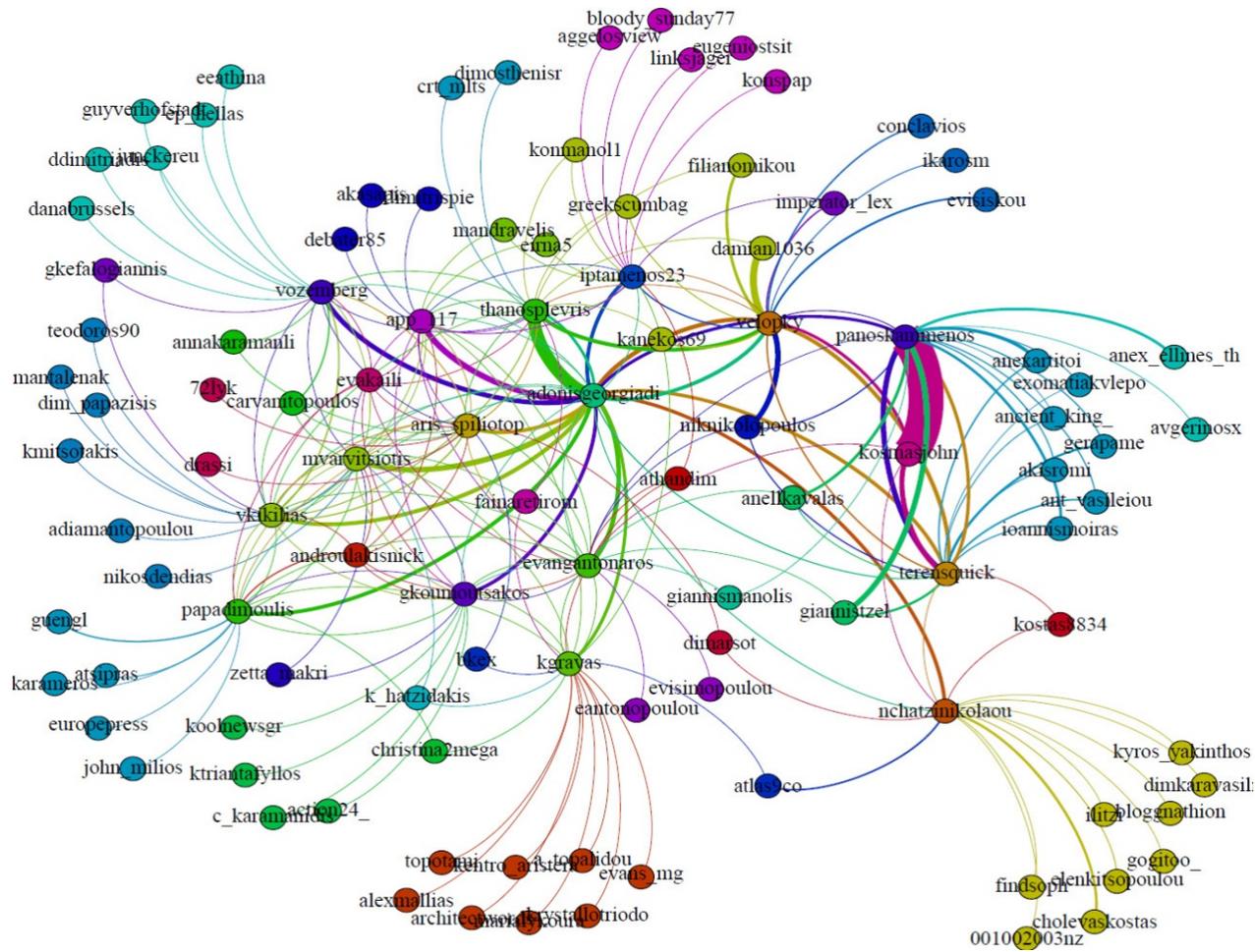

Figure 5: Case study similarity network (depth=2) – Thicker edges denote more similar Twitter accounts

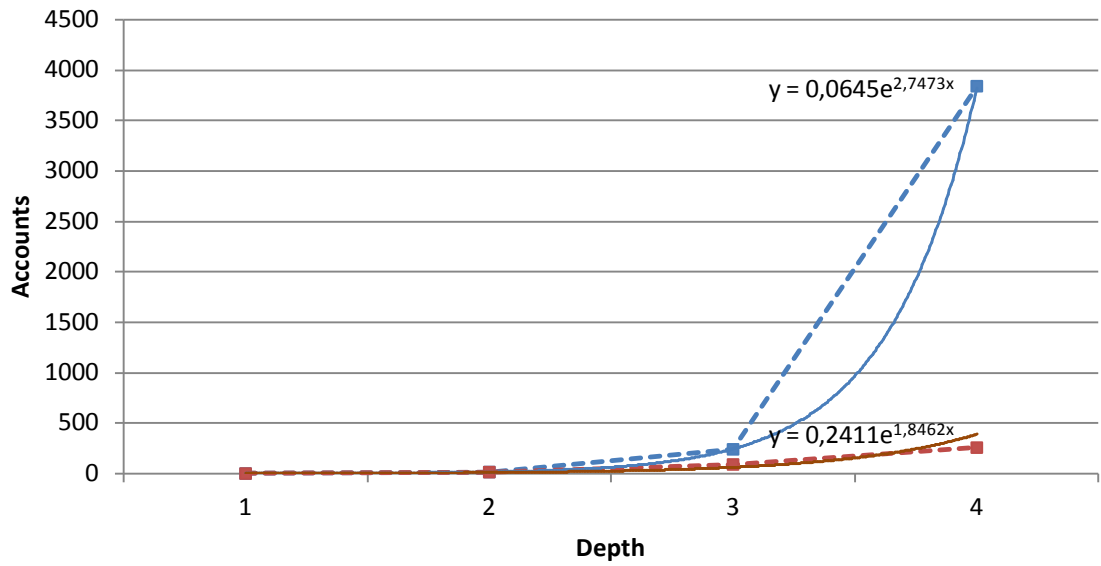

Figure 6: The theoretical maximum number of accounts (per depth) vs. the actual unique ones inserted into the network – Trending behavior is according to exponential type (values are depicted in Table 4)

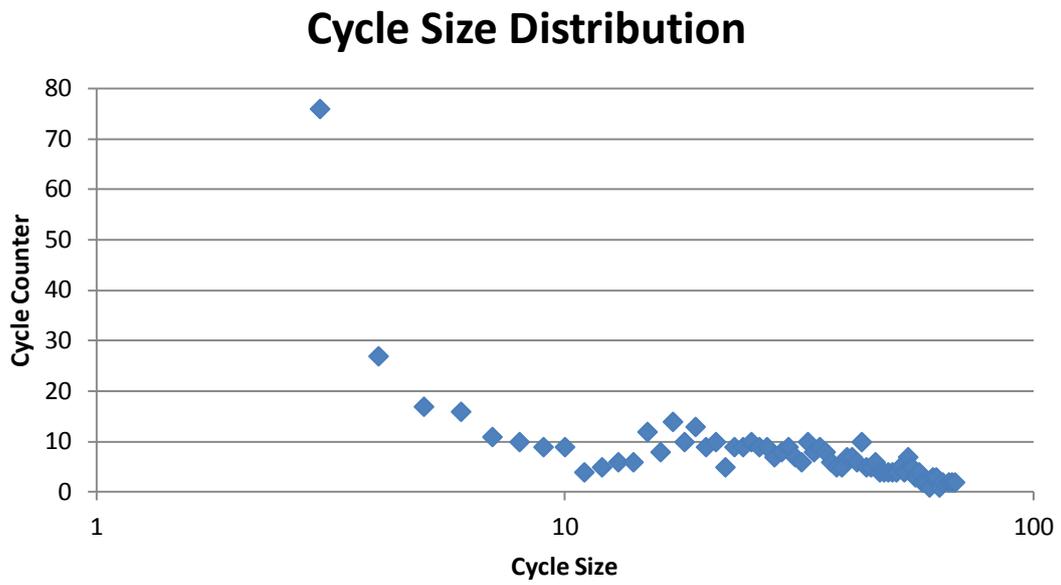

Figure 7: The closed cycle size distribution of a network (356 unique accounts – 531 closed cycles)

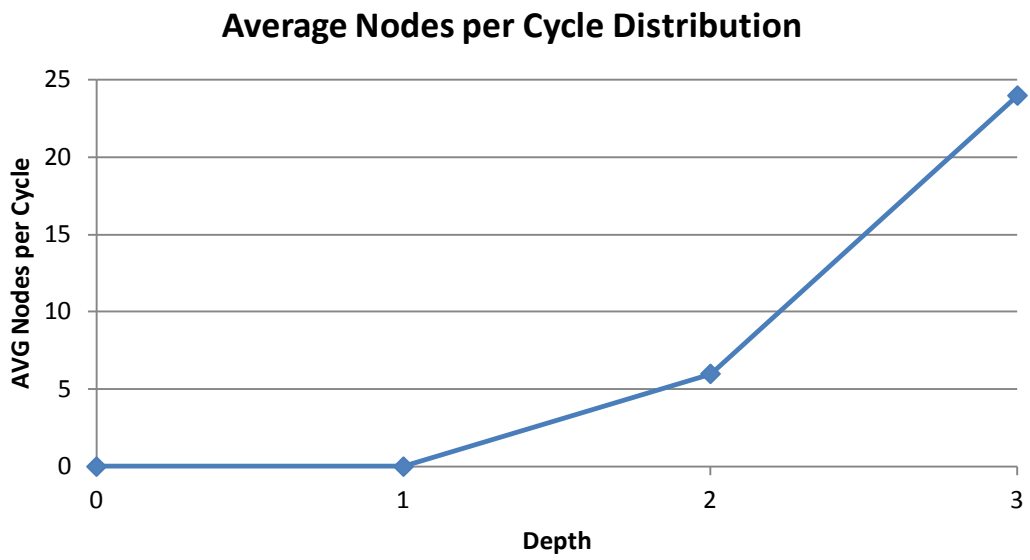

Figure 8: The average nodes per cycle distribution for each depth

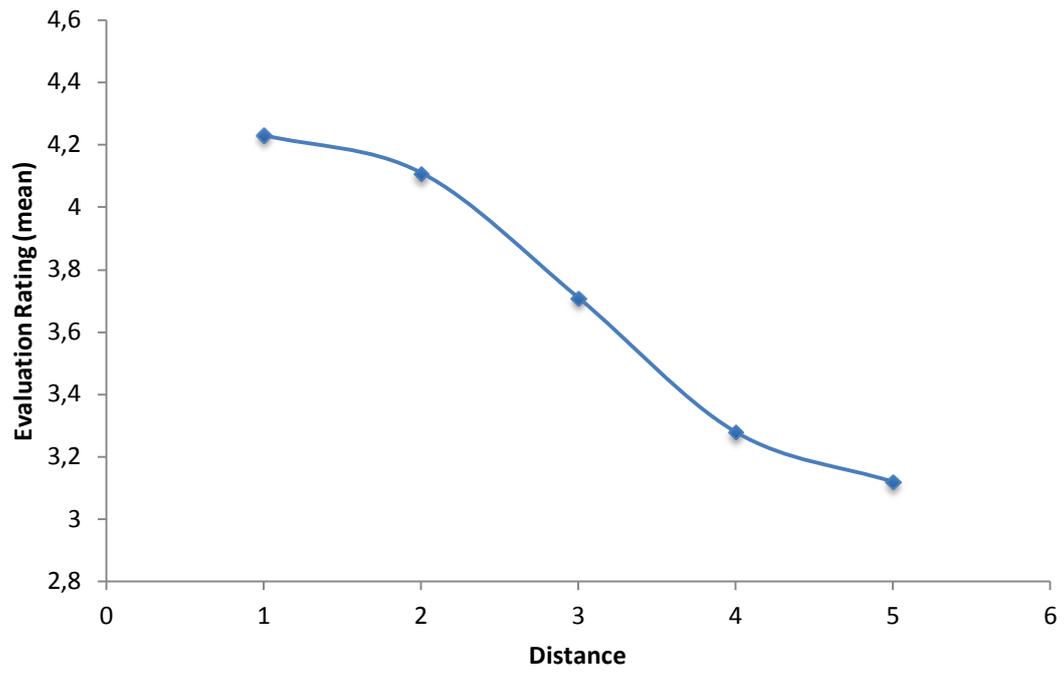

Figure 9: Mean rates (from the evaluators) vs. Distance in Similarity Network

## Table of Tables

| RT *h*-index | *h*h* | Transformed as | Calculation Process | Adjusted Tweets |
|---|---|---|---|---|
| 0,3 | - | 0,3 * 10^0 | 0,3/10, 10^0 | 0,03 |
| 2 | 4 | 4 * 10^0 | 4/10, 10^0 | 0,4 |
| 6 | 36 | 36 * 10^0 | 36/10, 10^0 | 03,6 |
| 15 | 225 | 22,5 * 10^1 | 22/10, 10^1 | 12,2 |
| 45 | 2.025 | 20,25 * 10^2 | 20/10, 10^2 | 22 |
| 80 | 6.400 | 64 * 10^2 | 64/10, 10^2 | 26,4 |
| 100 | 10.000 | 10 * 10^3 | 10/10, 10^3 | 31 |

Table 1: Calculating the "Adjusted Tweets"

| Accounts | 90,578 | URLs | 188,542 | Hashtags | 38,020 |
| Full Accounts | 2,423 | Full URLs | 72,931 | Images | 59,160 |
| Simple Accounts | 88,155 | Domains | 8,402 | Triples | 1,982,367 |

Table 2: The contents of the queried graph

| account | hashtags | | mentions | | URLs | | domains | | CWC | categories (out of 4) | SM |
|---|---|---|---|---|---|---|---|---|---|---|---|
| | total | common | total | common | total | common | total | common | | | |
| @thanosplevris | 3 | 2 | 98 | 50 | 130 | 9 | 5 | 5 | 9.848 | 4 | 9.848 |
| @app_117 | 18 | 5 | 59 | 29 | 52 | 0 | 23 | 18 | 6.303 | 3 | 4.727 |
| @mvarvitsiotis | 37 | 6 | 83 | 32 | 303 | 1 | 1 | 1 | 4.518 | 4 | 4.518 |
| @vozemberg | 40 | 3 | 65 | 26 | 93 | 3 | 2 | 2 | 3.913 | 4 | 3.913 |
| @velopky | 25 | 4 | 373 | 57 | 410 | 4 | 5 | 4 | 3.411 | 4 | 3.411 |
| @evangantonaros | 2 | 2 | 29 | 18 | 63 | 0 | 23 | 11 | 4.521 | 3 | 3.390 |
| @vkikilias | 59 | 4 | 74 | 29 | 309 | 0 | 3 | 3 | 4.292 | 3 | 3.219 |
| @papadimoulis | 55 | 5 | 128 | 38 | 402 | 0 | 2 | 1 | 4.067 | 3 | 3.050 |
| @aris_spiliotop | 20 | 3 | 37 | 19 | 65 | 0 | 21 | 12 | 4.052 | 3 | 3.039 |
| @gkoumoutsakos | 9 | 1 | 46 | 22 | 51 | 0 | 2 | 1 | 3.776 | 3 | 2.833 |
| @iptamenos23 | 10 | 2 | 71 | 27 | 45 | 0 | 1 | 1 | 3.746 | 3 | 2.810 |
| @terensquick | 34 | 2 | 313 | 57 | 923 | 0 | 4 | 1 | 3.706 | 3 | 2.780 |
| @nchatzinikolaou | 68 | 7 | 309 | 47 | 1175 | 6 | 5 | 3 | 2.743 | 4 | 2.743 |
| @panoskammenos | 0 | 0 | 134 | 35 | 249 | 3 | 9 | 4 | 3.404 | 3 | 2.553 |
| @kgravas | 18 | 1 | 16 | 12 | 88 | 0 | 2 | 1 | 3.234 | 3 | 2.425 |

Table 3. The top-15 similar accounts of @adonisgeorgiadi along with their respective similarity metrics for the selected case study

| Depth | Maximum Accounts | Increase |
|-------|------------------|----------|
| 0 | 1 | 1 |
| 1 | 16 | 15 |
| 2 | 241 | 91 |
| 3 | 3841 | 258 |

Table 4: The maximum number of accounts and the unique ones inserted into the network